\documentstyle[12pt]{article}
\input epsf
\newcommand{\svec}[1]{\mbox{\boldmath $#1$}}

\begin{document}
\title{\tt Reply to K. Amos et al nucl-th/0401055.}
\author{Jacques Raynal}
\date{  }
\maketitle
\centerline{\it  4 rue du Bief, 91380 Chilly--Mazarin, France}
\begin{abstract}
An expression for the spin--orbit interaction coupling between
different levels, which was shown to be aberrant more than thirty
years ago persists in the literature without clear indication of what
is used. It leads to expressions quite simpler than they should be.
After an attempt to warn the community of the nuclear physicists on
this strange situation (nucl-th/0312038), the authors of the
publication in which the ``aberrant'' interaction is described and
used, try to justify their work (nucl-th/0401055), by a very strange
``symmetrization'' of something already symmetric. They claim also
that their method allows to solve some problem related to the Pauli
principle and give some references, among which a book which reports
the solution of such problem almost forty years ago, with a very small
effect. An examination of their own results shows that their optimism
is not completely justified. Nevertheless, any user of {\tt ECIS},
sensitive to their arguments, is requested to ask their opinion to
these five coauthors before publishing.
\vskip .5truecm
PACS numbers 24.10.-i,25.40.Dn,25.40.Ny,28.20.Cz
\end{abstract}

After the publication of some article \cite{RAP1} in Nuclear Physic A,
I wanted to publish a comment \cite{RAP2} warning the nuclear physicist
community that two different deformed spin--orbit are used in the
literature for nucleon--nucleus inelastic scattering without notifying
which one is employed. Because everybody uses the same expression in
nuclear structure studies (for this reason, I qualified it as
``normal''), for inelastic scattering, I called ``normal'' the one which
has the same behavior between partial waves and ``aberrant'' the other
one. As I am not of native of English language, I was thinking to be
allowed to use this word, because I found on page 3 of \cite{RAP3}~:
\vskip -1.cm
\begin{quote}
{\bf ab--er--rant} ($\dots $),{\it adj.} $[\dots ]$, deviating from
what is true, correct, normal, or typical.
\end{quote}
and on page 3 of the first volume of \cite{RAP4}~:
\begin{quote}
{\bf ab--er--rant} $\backslash \dots \backslash $ {\it adj}
$[\dots ]$, {\bf 1 :} straying from the right or normal way~:
deviating from truth, rectitude, propriety~~~ {\bf 2 :} deviating
from the usual or natural type~~~ {\bf :}
{\scriptsize {EXCEPTIONAL, ABNORMAL. }}
\end{quote}
($\dots $ denoting pronunciation and etymology). My problem to
publish such an advertisement is that I have only \cite{RAP1} to cite.

On the 14th of June, I was visiting the office of the Nuclear Energy
Agency, at Issy--les--Moulineaux, which send nuclear codes to who wants
them all over the world (except in USA). We found on the web the answer
\cite{RAP5} of which the title is quite terrifying for some body
who sent almost 300 {\tt ECIS} to more than 50 countries. I find this
answer largely out of the subject. I have to answer to the spin--orbit
question, but also on the antisymmetrization with occupied states,
another subject on which I have comments to do but which I did not
want to publish.
\vskip .5truecm

In \cite{RAP2}, Eq. (1) is not the spin--orbit interaction but its
coefficient; the spin--orbit interaction involves the two following
lines. The six--parameter spin--orbit interaction of {\tt ECIS} can
be used to compare results with the two different deformed spin--orbit
interactions and many other expressions by who wants it; anyway, the
parameters are read only if a special logical is set true. This is like
that since {\tt ECIS67}. Instead of laughing at, why the authors of
\cite{RAP1} do not do the same?

As the authors of \cite{RAP1} say that the ``normal'' is {\it derived
in some extremely hard--to--find publications''}, let us resume it.
First of all, the two-body spin--orbit interaction as described in
\cite{RAP6,RAP7,RAP8,RAP9} {\footnote {In \cite{RAP6} there was an
error for $a^J(1,2)$ as said in \cite{RAP7}~: the coefficients of
$V_{J-1}$ and $V_{J+1}$ are $J(J-1)$ and $(J+1)(J+2)$ instead of
$J(J+3)$ and $(J+1)(J-2)$ respectively. There is also a factor 4
in the three first publications, coming from an error in writing the
derivative with respect to $(\svec r_1 - \svec r_2)$ and assimilation
of $\svec \sigma $ to $\svec s$. In \cite{RAP9},
$-(\alpha ^J_{j_2j'_2})^2$ and a factor $1/4$ are missing in the
expression of $d^J(1,2)$.}} is ``normal''; therefore,
the intermediate step of its use as a one-body interaction, described
by the sums of two terms of Eq. (84) of \cite{RAP6}, the Eqs. (21-22)
of \cite{RAP7}, the Eqs. (51-52) of \cite{RAP8}, and Eqs. (4,49-50)
of \cite{RAP9} are also ``normal''.  With a finite range, these
expressions are a kind of folding potential for the direct term, to
which must be added similar expressions for the exchange term. At the
zero--range limit (not $\delta (\svec r - \svec r')$ but $\delta
''(\svec r - \svec r')$), direct
and exchange terms for the two--body interaction are identical~:
they are given by Eq. (30) of \cite{RAP7} and Eq. (55) of \cite{RAP8}.
Assuming zero for the eigenvalues of $(\svec l \ . \ \svec \sigma )$ of
one particle (which means complete shells for $l=j+1/2$ and $j=l-1/2$
with same radial function), one get the same expression as with the
``full Thomas term''. The result must be made hermitian, but not in
the sense used by the authors of \cite{RAP1,RAP5} by dropping the
derivative term acting on the right side but by replacing it by a
derivation on the left side with opposite sign (that is~: acting also
on the form--factor).

As the ``normal'' coupled--channel spin--orbit potential is {\it
derived in some extremely hard--to--find publications} \cite{RAP5}
but cannot been published anywhere because it is not new, let us
give it here in details. First of all, the spin--orbit obtained when
the Dirac equation is changed into its Schr\"odinger equivalent
(``full Thomas form'') can be written as~:
\begin{equation}
V^{LS} = \sum_{\lambda ,\mu }\Big( \nabla V_{\lambda }(r)
Y^{\mu }_{\lambda } ({\hat r})\Big) \times {\nabla \over i}  \ . \
{\svec \sigma }  \label{E1}
\end{equation}
which avoids to deal with more equations than necessary~: the zeroth
order term of \cite{RAP1} is in $V_0(r)$, the
first order term for some $\beta _2$ is in $V_2(r)$, the second
order is $V_4(r)$ and also in $V_0(r)$ and in $V_2(r)$ (for the $n^{th}$
order, it is in all even $V$, from $V_0(r)$ to $V_{2n}(r)$). Using
the following identities~:
\begin{eqnarray}
\nabla ={\svec r \over r} \ {d \over {dr}} - i {{\svec r \times
\svec \ell } \over {r^2}},  \qquad \qquad i \svec \sigma .(\svec A
\times \svec B ) = (\svec \sigma .\svec A) (\svec \sigma . \svec B)
- (\svec A .\svec B),&& \nonumber \\ (\svec \sigma .\svec \nabla)=
{(\svec \sigma .\svec r) \over r} \Big({d \over {dr}}
- {{(\svec \ell .\svec \sigma )} \over r} \Big), \
(\svec \sigma .\svec \ell) (\svec \sigma .\svec r) = - (\svec \sigma .
\svec r) (\svec \sigma  .\svec \ell), \ (\svec \sigma  .\svec r)^2
= r^2,&& \label{E2}
\end{eqnarray}
$V^{LS}(r)$ can be written as~:
\begin{eqnarray}
V^{LS} &=& \sum_{\lambda ,\mu } - \Big( \Big[{d \over {dr}} +
{{(\svec \ell \ . \ \svec \sigma )}\over r} \Big] V_{\lambda }(r)
Y^{\mu }_{\lambda } ({\hat r}) \Big) \Big[{d \over {dr}} -
{{(\svec \ell \ . \ \svec \sigma )}\over r} \Big] \nonumber \\
&+& \Big({d \over {dr}} \ V_{\lambda }(r)
Y^{\mu }_{\lambda } \ ({\hat r}) \Big) \ {d \over {dr}} \ - \
\Big( {{\svec r \times \svec \ell } \over {r^2}} V_{\lambda }(r)
Y^{\mu }_{\lambda } ({\hat r}) \Big) \ {{\svec r \times \svec \ell }
\over {r^2}} \label{E3}
\end{eqnarray}
The terms with two derivatives cancel one another. Noting by $\ell _i$
and $\gamma _i$ the angular momentum and the eigenvalue of
$(\svec \ell .\svec \sigma )$ of the right side, $\ell _f$ and
$\gamma _f$ for the left side, $(\svec \ell .\svec \sigma )$
acting on $Y^{\mu }_{\lambda } ({\hat r})$ can be replaced by
$(\gamma _f -\gamma _i)$ because $\svec \ell _f = \svec \ell _i +
\svec \lambda $. The last term can be simplified, using the relation~:
\begin{equation}
(\svec A \times \svec B) . (\svec C \times \svec D) =
(\svec A . \svec C) (\svec B . \svec D)
- (\svec B . \svec C) (\svec A . \svec D) \label{E4}
\end{equation}
which replaces the two cross products by $r^2 (\svec \lambda .
\svec \ell _i)$. But, as $\svec \ell _f = \svec \ell _i + \lambda $
and $(\svec \ell . \svec \ell) = (\svec \ell . \svec \sigma )^2 +
(\svec \ell . \svec \sigma )$~:
\begin{equation}
- 2 (\svec \lambda . \svec \ell_i) = \lambda (\lambda +1) +
(\gamma _i - \gamma _f)(\gamma _i + \gamma _f +1) \label{E5}
\end{equation}
With these quite simple manipulations, the result is obtained as~:
\begin{eqnarray}
V^{LS} &=& \sum_{\lambda ,\mu } Y^{\mu }_{\lambda }(\hat r) \Big[
{{d V_{\lambda }(r)}\over {dr}} \ \gamma _i \ + \ {{V_{\lambda }(r)}
\over r} \ (\gamma _i - \gamma _f) {d \over {dr}} \nonumber \\ &+&
{{V_{\lambda }(r)}\over {2 r^2}} \ \Big\{ \lambda (\lambda +1) -
(\gamma _f - \gamma _i )(\gamma _f - \gamma _i \pm 1) \Big\}  \Big] 
 \label{E6}
\end{eqnarray}
where $\pm 1$ is $+ 1$ in this tri--dimensional derivation and is
$-1$ if the wave functions are multiplied by $r$ as usual. This
derivation should not be a problem to people used to angular momenta,
$\svec \sigma $--matrices, scalar and vector products.

To say that the first term of Eq. (\ref{E6}) is {\it fully consistent}
with the whole is quite strange. The fact that the two last terms can be
replaced \cite{RAP1,RAP5,RBP1} by a $(\svec \ell .\svec \ell )$ and
a $(\svec s .\svec s)$ interactions as yet to be proven. Note that
the first publications which used the ``full Thomas term''
\cite{RBP2,RBP3} did not notice the behavior $(\gamma _i - \gamma _f)$
of this term because they ignored the (quite simple) derivation
presented above. This interaction is expected to play a role primarily
for the asymmetry of the inelastic scattering, but less than the
deformed central interaction~: it should be so in the relation of the
amplitude of this asymmetry with the sign of the deformation in
the rotational model \cite{RBP4}. Anyway, If the deformed spin--orbit
interaction plays no role in their problem \cite{RAP1}, why they use it.

The symmetrization of an operator including $d/dr$ acting on the
right side is its replacement by $-d/dr$ acting on the left side,
that is on the form--factor as well as the function. The use of the
deformed spin--orbit \cite{RAP1} is equivalent to the use of~:
\begin{equation}
V^{LS}_{ijkl}(r)=V_{ijkl}(r) \{[\ell .\svec {s}]_i+[\ell .\svec {s}]_j
+[\ell .\svec {s}]_k+[\ell .\svec {s}]_l\} \label{E7}
\end{equation}
for the two--body spin--orbit interaction, as can be seen after one
integration.
\vskip .5truecm

In \cite{RAP5}, there are many comments and references related to the
{\it Pauli principle} of which it was not question in \cite{RAP2}.
I was allowed by the Service de Physique Theorique to photocopy all
the reference [2] of \cite{RAP5} in its library (including the
second one in Saclay's central library) and also the 3 references of
the article in Nuclear Physics related to Pauli principle (130 pages
for all that) and to borrow \cite{RBP5}. It seems that they never
opened this book; I did not remember of its content. There is a very
good table of references by which I found myself cited 8 times as
RA 67b, once as RA 68 and also twice as GI 67 and once as ME 66.
The first \cite{RBP6} of their references [2], of G. Pisent, one of
the coauthors, is also cited twice as PI 67c in this book~: a footnote
on page 103 ({\it In the papers concerned with ${}^{13}C$ as a
compound nucleus the exclusion principle could not be exactly
satisfied $\vert $ ... , ... , PI 67c, ... $\vert $}.), and the last
paragraph of page 113 ({\it ... and Pisent and Saruis $\vert $PI
67c$\vert $. This various works suffer from the drawback that the Pauli
exclusion principle is violated at some stage of the calculation.}).

In Spring 1965, I was theoretically at USC, practically at UCLA, in
Los Angeles. With M. A. Melkanoff and T. Sawada, we decided to do
some calculations on the giant resonance of ${}^{16}O$ using the shell
model with a continuum theory of C. Bloch and V. Gillet \cite{RBP7}.
This work has been published in Nuclear Physics A \cite{RBP8} and
as a seminar at a Summer School in Varenna \cite{RBP9}; in the same
book, there is another seminar from me on the "Stretch scheme" and a
seminar entitled {\it Results of Hartree--Fock calculations with
non--local and hard--core potentials}, by J. P. Svenne, Canadian of
Copenhagen University, who I think to be one of the coauthors of
Ken Amos. This work is partly reported in \cite{RBP5}. The space used
is described in the book on page 23 in the text together with figure
3.2 and its legend. C. Bloch and V. Gillet could obtain values only
at points which they choose for the grid, but we managed to obtain
continuous results (footnote on page 77~:{\it Care must be taken
because the integrands involve $a^c(E'';c)$ which is singular at
$E''=E$ $\vert $RA 67b}$\vert $.) with a minimum number of points and
did computation from 16 to 30 MeV (this is scattering on ${}^{15}O$ for
which 16 MeV is the threshold with respect to ${}^{16}O$). Then, we
decided to do the same calculation in the $r$--space instead of the
$E$--space. We got different results; looking why, we orthogonalized
with the $1s_{1/2}$ occupied bound-state, thinking that a small
mixture of this state give very important effects for
${}^{16}O(\gamma ,n)$, more than for the elastic scattering because this
result is the integral of the solution multiplied by $r$ and the hole
function. We obtained the same result as in $E$-representation. That
was the proof that these two approaches, mathematically equivalent are
numerically equivalent (discarding error or imprecision on one of them).
This is presented pages 106-110 with the results in figure 6.1 . On
page 103, {\it 6.3a. Coupled channels approach} the first paragraph
includes two citations prior to \cite{RBP8} as not taking into account
antisymmetrization and quote RA 67b as showing this effect. The second
paragraph is~: {\it The most complete coupled channels calculation of
the reactions ${}^{16}O(\gamma ,n)$ and ${}^{16}O(\gamma ,n)$ (for
$E1$ transitions) was carried through by Raynal, Melkanoff and
Sawada $\vert $RA 67b$\vert $. These authors treat
antisymmetrization correctly.} In fact, I never saw the {\it Pauli
principle} expressed more clearly than by Eqs. (19-20) of \cite{RBP9}
or Eqs. (55-58) of \cite{RBP8} (Eqs. (59-61) for a zero--range
interaction). The reference \cite{RCP1} given in \cite{RAP1} uses only
a $1s (\alpha )$ state with no generalization.

More details can be found in \cite{RBP8,RBP9}~: figures 12 and 13
in the first reproduced by figures 4 and 3 in the second show the
results obtained respectively for ${}^{16}O(\gamma ,n)$ and
${}^{16}O(\gamma ,p)$ with five channels and coupling the ten
channels. In neutron figures, there are~:
\begin{description}
\item {$\bullet $} the five channels result,
\item {$\bullet $} the ten channels one,
\item {$\bullet $} the experimental results known at that time,
\item {$\bullet $} and also results obtained with five channels
without taking into account the occupation of the $1s_{1/2}$ state.
\end{description}
Unhappily, this last curve is not given in figure 6.1 of \cite{RBP5}
which shows only five channels results and experimental data for
neutrons and not this fourth curve which is essential to clarify the
point in discussion: there is no noticeable effect up to 20 MeV
(4.5 MeV above threshold) but a shift of the maximum around 22 MeV,
about the same as between five and ten channels calculations.
In the same two publications, we showed that the difficulty to deal
with a resonance $d_{3/2}$ in the continuum in E--representation
can be overcome by using a bound--state and taking into account
the difference of the Saxon--Woods wells in r--representation or
E--representation. In \cite{RBP8,RCP2}, we studied the effect of
a 2p--2h state as quoted GI 67 by \cite{RBP5} on pages 108 and 225.

I foresee Ken Amos' answer: it is not the same problem, you deal with
${}^{15}O$ and ${}^{15}N$ and not ${}^{12}C$, you use some two--body
interaction instead of a pure one--body, and so on, and so on ... But
look to their own results, table I, page 86 of \cite{RAP1} the three
lines where there are experimental data and results with and without
OPP~: for the first, the energy is shifted by 33\% of what is needed,
in the good direction but the width is increased of 20\% only instead
of 148\%; in the second one the energy is shifted of only 4\% of what
is needed and the width unchanged; in the third one, the energy is
shifted of 82\% of what is needed, in the good direction (great
success) but the width is increased 15\% instead of being divided by
4,33; in a fourth case, there is no effect. With these values, they
claim that this antisymmetrization is absolutely necessary; with the
same values, I feel that it disturbs the results.

On the 25th of June 1975, I participated to the jury for the thesis of
J.-M. Normand \cite{RCP3} at Orsay with V. Gillet, P. Benoist--Gueutal,
M. Goldman and R. Arvieu as president. He showed the effect of the
Pauli principle at threshold energies \cite{RCP4}. He studied
scattering lengths and effective ranges of neutrons, which I think
quite sensitive to these effects, on ${}^{12}C$, ${}^{13}C$,
${}^{16}O$, ${}^{17}O$, ${}^{19}F$ and ${}^{40}Ca$. Only the scattering
lengths were known at that time. In table 5, using 4 different
interactions with different strengths (in all 14 calculations), he
found for ${}^{12}C$ a decrease of 8\% to 21\% for the scattering
length, of 4\% to 7\% for the effective range (but, among these 14
calculations, the smallest value is 54\% of the largest one for the
scattering length and 30\% for the effective range, 30\% and 20\%
discarding the largest value). In table 6, for ${}^{16}C$, also with
4 interactions and 13 calculations, these figures reduce to 5\% to
7\% for the scattering length, 2\% to 4\% for the effective range and
values spread by 12\% for both. In Table 7 for ${}^{40}Ca$, with 2
interactions and 6 calculations, there is no effect of the $1s_{1/2}$
state (less than 0.3\%) but a large effect of the $2s_{1/2}$, 21\%
to 42\% for the scattering length, 15\% to 31\% for effective range,
leading to almost identical results; the variation of this last results
are 0.6\% and 1.1\% for a variation of 4.2\% and 3.3\% of the strength
of the interaction (very special case for which the Pauli principle
is more important than the model, id est, than the strength of the
interaction). In table 8 are given 6 results for ${}^{13}C$ and 2 for
${}^{17}O$~; in this table, there are results for two values of $J$.
In all these cases, one can see that the corrected results are quite
near the uncorrected ones obtained with an increase of the interaction
by about 3\%. Results for ${}^{19}F$ show the importance of the choice
of the space of configuration.

Even if the effects are more important for ${}^{12}C$ than ${}^{16}O$,
I do not see in this very sensitive calculation a justification of
the assertion of Ken Amos that the Pauli principle affects strongly
results at low energy and it is not their publication which can convince
looking their table 1. Anyway, such phenomena are weaker with a complex
potential (because the wave function is damped inside the nucleus) and
{\tt ECIS} was not written for such problems. As said in the title,
any user of these codes who has the smallest doubt about this subject
should ask their opinion to K. Amos et al. Anyway, it is a lot easier to
add that to {\tt ECIS} than to introduce a quadratic spin-orbit
which was never seriously used in {\tt DWBA90}.

In \cite{RAP5}, these is a reference to page 426 of Hodgson's book 
\cite{RCP5}; in the following pages, there is a presentation of
\cite{RBP1} and of some publications of G. Pisent, before or after
\cite{RBP6} with no allusion to the ``Pauli exclusion principle''.
In the subject index, seven pages are indicated for this topic~:
\begin{description}
\item {$\bullet $} page 90 on semiclassical optical model,
\item {$\bullet $} page 113 for application in nuclear medium,
\item {$\bullet $} pages 130, 131, 132 for the calculation of the
imaginary part of the optical potential,
\item {$\bullet $} page 162 related to consequences for nuclear medium,
\item {$\bullet $} page 581 consider the effect of the excess of
neutrons on the difference between the number of reactions $(p,n)$
and $(n,p)$.
\end{description}
The ``Pauli exclusion principle'' applied to the scattering wave is
completely ignored in this book where this effect is only applied
to nucleon--nucleon scattering in nuclear medium as needed in
\cite{RCP6}~: the critics of \cite{RAP5} on the conception of {\tt
ECIS} are also valid for it. There is no question of spin--orbit
deformation~: \cite{RBP2,RBP3} are not cited. My own thesis is cited for
different points (an error on page 148, not reproduced on pages 231,
235 and 244) including figure 10.6 on elastic deuteron scattering.

The third book \cite{RCP7} cited in \cite{RAP5} is not available at
Saclay. I cannot afford to buy it and analyze it as I did above for the
two first ones.

I hope that every user of {\tt ECIS} will make his own opinion on
\cite{RAP5} and my answer, even those who use it for heavy--ion
scattering because the title does not exclude this subject. If they
have any doubt, they should communicate their results to the five
coauthors of which they can find the e--mail address in \cite{RAP1}.
\vskip .5truecm

{\bf Conclusions}

\begin{description}
\item {$\bullet $} When they say~: {\it the spin--orbit expression we
use is fully consistent with the $\svec S . \svec L$ term that comes
from the full--fledged Thomas term}, they forget to add that they
pluck.
\item {$\bullet $} If they open the books which they give as reference,
they can see that the effects which they claim to be at low energy
are seen only at higher energy. Even if they are some effects
\cite{RCP3,RCP4}, they can disappear if there is a search on
parameters as in \cite{RAP1}.
\item {$\bullet $} If they look at the table which they publish, they
have to agree that the Pauli principle is inefficient to give good
results~: readers can conclude that their method is bad or that the
Pauli principle has no notable effects but cannot agree with their
optimistic comments.
\end{description}
Anyway, the Pauli principle was not the subject of \cite{RAP2} but
the fact to see in the literature an expression which I believed
forgotten since a long time and was certainly used many times without
being quoted. The ``normal'' expression is easy to derive as shown
by equations (\ref{E2}) to (\ref{E6}) there in. The allusion to a
mosquito and an elephant at the end of \cite{RAP5} reminds a tale
of La Fontaine about a frog and an ox. {\it Errare humanum est}
(see footnote); {\it perseverare diabolicum}.

\end{document}